\documentclass{article}
\usepackage{spconf,amsmath,amssymb,graphicx,hyperref}
\usepackage{booktabs}

\usepackage{multirow}
\usepackage{array}
\usepackage{xcolor}
\usepackage{colortbl}
\usepackage{makecell}
\usepackage{balance}
\usepackage{afterpage}
\usepackage{tikz}
\usetikzlibrary{arrows.meta,positioning,shapes.geometric,calc}

\newcommand{\clnew}[1]{#1}

\newcommand{\rev}[1]{#1}

\title{The Vulnerability of Neural Audio Watermarks under
       Speech Enhancement}

\name{Xincong Zhong$^1$, Shengyao Wang$^1$, Lingfeng Yao$^2$, Yihang Bao$^1$, Jinze Yu$^1$, Miao Pan$^2$, Jiang Liu$^{1}$\sthanks{Corresponding author.}}
\address{$^1$Waseda University, Japan \\
         $^2$University of Houston, USA}

\begin{document}
\ninept
\setlength{\parskip}{0pt}
\maketitle

\begin{abstract}
Neural audio watermarks are increasingly deployed in commercial
speech generation systems to make AI-generated speech traceable, yet
their robustness has been studied mainly under conventional signal
distortions. Since a watermark can be
regarded as imperceptible noise added to the speech signal, a natural
question is whether speech enhancement (SE), as a denoising model, can
remove it. In this paper, we cascade Gaussian noise with SE models as
a black-box watermark removal attack, covering both discriminative and
generative SE paradigms, against six neural watermarks: AudioSeal,
WavMark, SilentCipher, Timbre, Perth, and AlignMark. Experimental
results show that the proposed attack significantly outperforms
existing neural re-synthesis methods in watermark removal. In
particular, we find that generative SE, which reconstructs the harmonic
regions of speech while denoising, is highly destructive to
watermarks. \rev{These findings show that SE poses a serious threat to
current audio watermarking methods, and we call for SE-aware robustness
evaluation in watermark design.}
\end{abstract}


\section{Introduction}
\label{sec:intro}

\rev{The rapid development of speech generation technology has
substantially narrowed the perceptual gap between synthetic and human
speech~\cite{xu2025qwen3}, thereby facilitating impersonation, fraud,
and misinformation. Passive deepfake detectors~\cite{xie2023domain} can indicate whether
speech is AI-generated, but they cannot reliably trace it back to its
source. This limitation motivates proactive provenance
mechanisms that mark speech when it is generated.
Neural audio watermarking is one such mechanism, with
imperceptible signatures embedded into generated
speech~\cite{liu2024audiomarkbench}. Recent
methods~\cite{roman2024proactive,chen2023wavmark,singh2024silentcipher,huang2024timbrewatermarking,liu2025xattnmark,li2026alignmark}
report strong robustness while maintaining high perceptual quality,
and watermarking has been integrated into practical speech generation
systems~\cite{resembleai2024perth}.}

\rev{The reported robustness has been measured mainly against
conventional signal-level transformations, including additive noise,
lossy compression, filtering, resampling, pitch shifting, and time
stretching. Beyond these transformations, several studies propose
intentional removal attacks. AudioMarkBench formulates watermark
removal as constrained optimization and uses gradient-based
perturbations to reduce watermark
detectability~\cite{liu2024audiomarkbench}. Overwriting attacks inject
forged watermarks to interfere with or replace legitimate
ones~\cite{yao2025yours}, while learning-based approaches such as
HarmonicAttack train dedicated removal models from watermarked and
clean speech pairs~\cite{li2025harmonicattack}. Diffusion-based
mel-spectrogram reconstruction followed by neural vocoding has also
been shown to remove watermarks~\cite{yao2026audio}.
Even neural re-synthesis through codecs and vocoders can fully remove
some neural watermarks~\cite{oreilly2025deep}.}

\rev{From the perspective of speech enhancement (SE), a neural watermark
is essentially imperceptible noise added to clean speech, which SE is
designed to remove. Modern SE not only denoises but also increasingly
reconstructs speech with generative
models~\cite{shetu2026comparison}, so a watermark must
withstand both. However, SE is rarely considered in the
robustness evaluation of existing watermarking schemes, and few attack
studies exploit it: DNN-based SE removes speech watermarks once trained
on paired clean and watermarked
speech~\cite{lopezlopez24_iberspeech}, and noise injection followed by
a discrete-token denoiser substantially suppresses watermark
detection~\cite{oreilly2025deep}.}

In this paper, we present the first systematic paradigm-level comparison of
SE-based neural audio watermark removal. In
particular, we organize SE into discriminative and generative
frameworks, and evaluate an attack pipeline that adds Gaussian noise
before SE against six neural watermarks under strict black-box
conditions. Results show that language-model-based generative SE
achieves the strongest overall watermark removal. Visualizations
further indicate that neural watermarks are particularly vulnerable to
the reconstruction of speech harmonic regions during generative
denoising. The remainder of this paper is organized as follows.
Section~\ref{sec:method} describes the evaluated watermarks, SE
models, and attack pipeline. Section~\ref{sec:eval} presents the
experimental results and analyzes the vulnerabilities exposed by these
attacks. Conclusions are summarized in Section~\ref{sec:conc}.

\section{Methods}
\label{sec:method}

\subsection{\rev{Audio Watermarking Methods}}
\label{ssec:watermarks}
We evaluate \clnew{six} neural watermarks: \clnew{five} state-of-the-art research
schemes and one closed-source industrial watermark.

\textbf{AudioSeal}~\cite{roman2024proactive} embeds the watermark as a
residual waveform perturbation produced by an EnCodec-style neural
encoder, jointly trained with a detector that outputs a sample-level
localization score.

\textbf{WavMark}~\cite{chen2023wavmark} embeds a message (a
synchronization pattern plus a payload) as a residual waveform
perturbation produced by an invertible convolutional network whose
weights are shared with the detector. Detection performs a
brute-force temporal shift to lock onto the synchronization pattern
and reports the bit accuracy of the recovered message.

\textbf{SilentCipher}~\cite{singh2024silentcipher} embeds a message
into mid-frequency bands via a CNN encoder/decoder; \rev{a signal-to-distortion
ratio (SDR)} controller and a psychoacoustic model
jointly hold the perturbation below the perceptual masking
threshold of the host audio.

\textbf{\rev{Timbre}}~\cite{huang2024timbrewatermarking} embeds
a message via a CNN on the magnitude spectrogram and recovers the
waveform by iSTFT with the original phase. A training-time
distortion layer simulating Griffin-Lim resynthesis hardens it
against vocoder-based voice-cloning regeneration attacks.

\textbf{Perth}~\cite{resembleai2024perth} is a neural watermark
deployed by default in a widely used commercial \rev{text-to-speech} system
(Chatterbox by Resemble AI). Its encoder places payload energy in
frequency bands chosen by a psychoacoustic model so that the
perturbation lies below the human auditory masking threshold.

\clnew{\textbf{AlignMark}~\cite{li2026alignmark} is a feature-aligned
watermark: it aligns the embedded signal with the original speech
content feature distribution, which lets it carry higher watermark energy to
improve robustness against neural re-synthesis attacks.}

\subsection{Speech Enhancement Methods}
\label{ssec:se-methods}
\rev{We evaluate six SE models from two paradigms, discriminative and
generative, for watermark removal. The two paradigms differ in how they
recover clean speech~\cite{shetu2026comparison}. Discriminative SE
treats enhancement as a regression problem and learns a deterministic
mapping from noisy to clean speech by minimizing a signal-level or
mask-level loss. Generative SE instead learns the distribution of clean
speech and generates the enhanced speech conditioned on the noisy
input.}

\textbf{Discriminative SE.} Demucs~\cite{defossez2020real}
is a causal encoder--decoder operating directly on the waveform.
DeepFilterNet3 (DF3)~\cite{schroter23b_interspeech} performs deep
filtering on equivalent-rectangular-bandwidth sub-bands and
predicts a complex mask.

\clnew{\textbf{Generative SE} further splits into
spectrogram-domain generative SE (STSE) and
language-model-based generative SE (LMSE).}
STSE operates in the continuous spectrogram domain. \clnew{We use
two models sharing the same network backbone and differing only in
the training objective:}
SGMSE~\cite{richter2023speech} learns a score
function under a stochastic differential equation (SDE) and
integrates the reverse diffusion with $N{=}60$ predictor--corrector
steps;
FlowSE~\cite{lee2025flowse} learns a vector field via flow
matching and integrates a deterministic ordinary differential
equation (ODE) with $N{=}10$ steps.
Instead of directly estimating a waveform or continuous
spectrogram, LMSE autoregressively predicts clean speech tokens, which
are subsequently decoded into a waveform by a neural codec.
Specifically, GenSE~\cite{yao2025gense} hierarchically generates clean
semantic and acoustic tokens, whereas UniSE~\cite{yan2025unise} predicts
a speech codec's global and semantic tokens from WavLM features and a
mel stream.

\subsection{\rev{Attack Pipeline}}
\label{ssec:attack-design}
We adopt a strict black-box attack model: the
attacker has access only to a single watermarked audio file,
with no knowledge of the watermark scheme, and aims only to
render the watermark undetectable while preserving perceptual
quality.

The attack has two steps. We first add Gaussian noise \rev{(GN)} to the
watermarked audio at a target input SNR of $20$ or $10$\,dB, then
enhance the noisy signal with one of the pretrained SE models
described in Section~\ref{ssec:se-methods}, using
the publicly released weights and the inference settings specified
there.

\section{Evaluation and Analysis}
\label{sec:eval}

\subsection{Datasets}
\label{ssec:dataset}
\rev{LibriSpeech~\cite{panayotov2015librispeech} is a corpus of about
$1000$ hours of read English speech derived from audiobooks and
sampled at $16$\,kHz.
Following~\cite{liu2024audiomarkbench,oreilly2025deep}, we select from
it a gender-balanced subset of $n{=}200$ utterances longer than
$5$\,s.}

\subsection{Baselines}
\label{ssec:baselines}
\clnew{%
We benchmark our SE-based attack against three neural
re-synthesis baselines: a codec round-trip via
EnCodec~\cite{defossez2022high} at $24$\,kbps; a mel-vocoder
round-trip via BigVGAN\,v2~\cite{lee2023bigvgan};
and audio-to-audio latent diffusion via
AudioLDM~\cite{liu2023audioldm} at transfer strength
$s{=}0.1$. Prior
studies~\cite{liu2024audiomarkbench,roman2024proactive,chen2023wavmark,huang2024timbrewatermarking,li2026alignmark,oreilly2025deep}
report these watermarks to be robust to classical signal-level
distortions such as filtering, downsampling and cropping, which we
therefore omit for space.}%

\subsection{Evaluation Metrics}
\label{ssec:metrics}
For watermark removal, following~\cite{liu2024audiomarkbench,oreilly2025deep},
we measure the true positive rate (TPR) at a $1\%$ false positive rate
(FPR), denoted \textbf{TPR@1\%FPR}, from detection scores (the bit
accuracy of the decoded message) over a balanced set of $200$ attacked
watermarked and $200$ unwatermarked LibriSpeech utterances. In
particular, AudioSeal and Perth are equipped with zero-bit detectors whose output is
thresholded at $0.5$; we therefore calibrate to $1\%$ FPR with this
threshold as a floor. In this paper, a watermark is considered
successfully removed when its TPR falls to $10\%$ or below.

To assess perceptual quality after attack, we use both
no-reference and reference-based metrics.

\textbf{No-reference perceptual metrics} estimate
speech quality from the attacked speech alone, with no access to a
clean reference.
\mbox{\textbf{UTMOS}}~\cite{saeki22c_interspeech} (UTokyo--SaruLab
Mean Opinion Score) is a universal regressor trained on listener ratings that
predicts speech quality on a scale of $1$ to $5$.

\textbf{Reference-based perceptual metrics}
compare the attacked speech with the original unwatermarked speech.
\textbf{ESTOI}~\cite{jensen2016algorithm} (Extended
Short-Time Objective Intelligibility) and
\textbf{ViSQOL}~\cite{chinen2020visqol} (Virtual Speech
Quality Objective Listener) are used to evaluate intelligibility
and perceptual quality, respectively.

\begin{table*}[t!]
\centering
\caption{Evaluation results of each watermarking method (first column) under the re-synthesis baselines and the proposed SE-based attack pipeline. Lower TPR@1\%FPR and higher UTMOS, ESTOI, ViSQOL indicate stronger removal and better preserved quality.}
\label{tab:main}
\scriptsize
\setlength{\tabcolsep}{1pt}
\renewcommand{\arraystretch}{1.25}
\begin{tabular*}{\textwidth}{@{\extracolsep{\fill}}c l |c|ccc|cccccc|ccccccc|ccccccc}
\toprule
\multicolumn{2}{c}{} & \multicolumn{1}{c}{} & \multicolumn{3}{c}{\textbf{Re-synthesis}} & \multicolumn{6}{c}{\textbf{SE only}} & \multicolumn{7}{c}{\textbf{GN 20\,dB + SE}} & \multicolumn{7}{c}{\textbf{GN 10\,dB + SE}} \\
\cmidrule(lr){4-6} \cmidrule(lr){7-12} \cmidrule(lr){13-19} \cmidrule(lr){20-26}
\textbf{WM} & \textbf{Metric} & \rotatebox{90}{WM only} & \rotatebox{90}{EnCodec} & \rotatebox{90}{BigVGAN\,v2} & \rotatebox{90}{AudioLDM} & \rotatebox{90}{Demucs} & \rotatebox{90}{DF3} & \rotatebox{90}{FlowSE} & \rotatebox{90}{SGMSE} & \rotatebox{90}{GenSE} & \rotatebox{90}{UniSE} & \rotatebox{90}{GN 20\,dB} & \rotatebox{90}{+\,Demucs} & \rotatebox{90}{+\,DF3} & \rotatebox{90}{+\,FlowSE} & \rotatebox{90}{+\,SGMSE} & \rotatebox{90}{+\,GenSE} & \rotatebox{90}{+\,UniSE} & \rotatebox{90}{GN 10\,dB} & \rotatebox{90}{+\,Demucs} & \rotatebox{90}{+\,DF3} & \rotatebox{90}{+\,FlowSE} & \rotatebox{90}{+\,SGMSE} & \rotatebox{90}{+\,GenSE} & \rotatebox{90}{+\,UniSE}  \\
\midrule
\multirow{4}{*}{\rotatebox{90}{\textbf{\makecell{AudioSeal}}}} & TPR@1\%FPR & 100.0 & 100.0 & \textbf{0.0} & \textbf{0.0} & 100.0 & 100.0 & 85.0 & 88.5 & \textbf{0.0} & \textbf{0.0} & 60.0 & 64.0 & 67.5 & 3.0 & 1.0 & \textbf{0.0} & \textbf{0.0} & 1.0 & 6.0 & 5.0 & \textbf{0.0} & \textbf{0.0} & \textbf{0.0} & \textbf{0.0}  \\
 & UTMOS & 4.09 & 3.75 & \textbf{3.99} & 2.64 & 4.12 & 4.14 & \textbf{4.17} & 4.16 & 3.38 & 4.07 & 3.06 & 3.87 & 4.00 & \textbf{4.08} & 4.08 & 3.42 & 4.05 & 1.49 & 3.52 & 3.59 & 3.82 & 3.85 & 3.27 & \textbf{4.06}  \\
 & ESTOI & 0.994 & 0.943 & \textbf{0.975} & 0.757 & 0.979 & \textbf{0.985} & 0.955 & 0.963 & 0.749 & 0.817 & 0.885 & 0.917 & \textbf{0.923} & 0.910 & 0.907 & 0.736 & 0.807 & 0.709 & 0.833 & \textbf{0.837} & 0.821 & 0.817 & 0.687 & 0.777 \\
 & ViSQOL & 4.941 & 4.567 & \textbf{4.899} & 4.175 & \textbf{4.730} & 4.685 & 4.493 & 4.583 & 3.755 & 3.967 & 3.801 & \textbf{4.057} & 3.930 & 3.953 & 3.943 & 3.442 & 3.761 & 2.867 & 3.518 & 3.425 & 3.431 & 3.376 & 3.094 & \textbf{3.562} \\
\midrule
\multirow{4}{*}{\rotatebox{90}{\textbf{\makecell{WavMark}}}} & TPR@1\%FPR & 100.0 & \textbf{0.0} & \textbf{0.0} & \textbf{0.0} & 100.0 & 100.0 & 99.5 & 100.0 & \textbf{0.0} & \textbf{0.0} & 28.0 & 6.0 & 9.5 & \textbf{0.0} & \textbf{0.0} & \textbf{0.0} & \textbf{0.0} & \textbf{0.0} & \textbf{0.0} & \textbf{0.0} & \textbf{0.0} & \textbf{0.0} & \textbf{0.0} & \textbf{0.0}  \\
 & UTMOS & 4.03 & 3.71 & \textbf{3.95} & 2.59 & 4.10 & 4.11 & \textbf{4.15} & 4.11 & 3.28 & 4.07 & 3.08 & 3.89 & 4.01 & \textbf{4.08} & 4.08 & 3.46 & 4.06 & 1.50 & 3.53 & 3.60 & 3.82 & 3.86 & 3.25 & \textbf{4.06}  \\
 & ESTOI & 0.988 & 0.941 & \textbf{0.969} & 0.753 & 0.975 & \textbf{0.979} & 0.948 & 0.961 & 0.738 & 0.817 & 0.889 & 0.919 & \textbf{0.926} & 0.913 & 0.910 & 0.738 & 0.806 & 0.712 & 0.835 & \textbf{0.839} & 0.823 & 0.819 & 0.684 & 0.780 \\
 & ViSQOL & 4.765 & 4.515 & \textbf{4.727} & 4.126 & 4.620 & \textbf{4.628} & 4.375 & 4.494 & 3.633 & 3.946 & 3.827 & \textbf{4.073} & 3.957 & 3.973 & 3.956 & 3.453 & 3.765 & 2.884 & 3.532 & 3.439 & 3.446 & 3.381 & 3.087 & \textbf{3.575} \\
\midrule
\multirow{4}{*}{\rotatebox{90}{\textbf{\makecell{Silent\\Cipher}}}} & TPR@1\%FPR & 99.5 & 1.5 & 5.0 & \textbf{1.0} & 99.5 & 99.5 & 95.0 & 97.5 & 1.0 & \textbf{0.5} & 1.0 & 1.5 & \textbf{0.5} & 3.0 & 1.0 & 1.5 & \textbf{0.5} & 1.0 & 2.0 & 1.5 & \textbf{0.0} & 0.5 & 1.0 & 0.5  \\
 & UTMOS & 4.10 & 3.77 & \textbf{4.02} & 2.65 & 4.13 & 4.15 & \textbf{4.18} & 4.17 & 3.29 & 4.11 & 3.07 & 3.89 & 4.01 & \textbf{4.09} & 4.08 & 3.39 & 4.06 & 1.49 & 3.52 & 3.60 & 3.82 & 3.86 & 3.27 & \textbf{4.07}  \\
 & ESTOI & 0.998 & 0.944 & \textbf{0.978} & 0.760 & 0.983 & \textbf{0.989} & 0.958 & 0.966 & 0.745 & 0.821 & 0.887 & 0.918 & \textbf{0.925} & 0.912 & 0.909 & 0.734 & 0.808 & 0.711 & 0.834 & \textbf{0.838} & 0.824 & 0.818 & 0.688 & 0.781 \\
 & ViSQOL & 4.938 & 4.539 & \textbf{4.894} & 4.247 & \textbf{4.728} & 4.698 & 4.495 & 4.581 & 3.715 & 3.997 & 3.819 & \textbf{4.068} & 3.946 & 3.965 & 3.951 & 3.435 & 3.763 & 2.875 & 3.525 & 3.435 & 3.441 & 3.385 & 3.097 & \textbf{3.573} \\
\midrule
\multirow{4}{*}{\rotatebox{90}{\textbf{\makecell{Timbre}}}} & TPR@1\%FPR & 100.0 & \textbf{0.0} & 100.0 & \textbf{0.0} & 100.0 & 100.0 & 100.0 & 100.0 & \textbf{0.0} & 0.5 & 98.5 & 91.0 & 86.0 & 37.0 & 19.0 & \textbf{0.0} & 0.5 & 46.5 & 15.5 & 27.0 & 2.5 & 1.0 & \textbf{0.0} & 0.5  \\
 & UTMOS & 3.97 & 3.65 & \textbf{3.87} & 2.56 & 4.02 & 4.06 & \textbf{4.12} & 4.08 & 3.26 & 4.01 & 3.03 & 3.86 & 3.98 & \textbf{4.07} & 4.06 & 3.42 & 4.04 & 1.48 & 3.49 & 3.56 & 3.79 & 3.83 & 3.25 & \textbf{4.06}  \\
 & ESTOI & 0.982 & 0.934 & \textbf{0.963} & 0.752 & 0.971 & \textbf{0.973} & 0.948 & 0.955 & 0.734 & 0.815 & 0.878 & 0.912 & \textbf{0.917} & 0.907 & 0.903 & 0.736 & 0.804 & 0.701 & 0.829 & \textbf{0.831} & 0.817 & 0.813 & 0.683 & 0.777 \\
 & ViSQOL & 4.698 & 4.433 & \textbf{4.662} & 4.085 & \textbf{4.564} & 4.536 & 4.384 & 4.448 & 3.624 & 3.920 & 3.714 & \textbf{4.009} & 3.871 & 3.910 & 3.895 & 3.427 & 3.739 & 2.803 & 3.484 & 3.379 & 3.397 & 3.343 & 3.062 & \textbf{3.553} \\
\midrule
\multirow{4}{*}{\rotatebox{90}{\textbf{\makecell{Perth}}}} & TPR@1\%FPR & 100.0 & 60.0 & 100.0 & \textbf{15.0} & 100.0 & 100.0 & 90.0 & 93.0 & \textbf{0.0} & 0.5 & 86.5 & 54.5 & 87.0 & 4.0 & 3.0 & \textbf{0.0} & 0.5 & 5.5 & \textbf{0.0} & 29.0 & \textbf{0.0} & \textbf{0.0} & \textbf{0.0} & \textbf{0.0}  \\
 & UTMOS & 4.05 & 3.70 & \textbf{3.96} & 2.59 & 4.08 & 4.10 & \textbf{4.14} & 4.13 & 3.33 & 4.08 & 2.98 & 3.82 & 3.95 & \textbf{4.03} & 4.03 & 3.44 & 4.03 & 1.47 & 3.45 & 3.52 & 3.76 & 3.79 & 3.22 & \textbf{4.05}  \\
 & ESTOI & 0.987 & 0.936 & \textbf{0.968} & 0.754 & 0.972 & \textbf{0.977} & 0.944 & 0.953 & 0.748 & 0.818 & 0.877 & 0.908 & \textbf{0.915} & 0.901 & 0.898 & 0.737 & 0.804 & 0.701 & 0.824 & \textbf{0.828} & 0.813 & 0.809 & 0.680 & 0.776 \\
 & ViSQOL & 4.927 & 4.542 & \textbf{4.886} & 4.184 & \textbf{4.713} & 4.674 & 4.469 & 4.558 & 3.735 & 3.992 & 3.779 & \textbf{4.036} & 3.909 & 3.929 & 3.915 & 3.438 & 3.758 & 2.840 & 3.492 & 3.403 & 3.401 & 3.346 & 3.067 & \textbf{3.551} \\
\midrule
\multirow{4}{*}{\rotatebox{90}{\textbf{\makecell{AlignMark}}}} & TPR@1\%FPR & 99.5 & 97.5 & 91.0 & \textbf{64.5} & 88.5 & 98.5 & 87.0 & 86.0 & 28.0 & \textbf{9.5} & 99.5 & 73.0 & 99.5 & 96.5 & 96.0 & 23.5 & \textbf{9.5} & 98.5 & 53.0 & 95.0 & 90.5 & 84.5 & 8.0 & \textbf{6.5}  \\
 & UTMOS & 3.79 & 3.45 & \textbf{3.71} & 2.48 & 3.81 & 3.83 & 3.88 & 3.86 & 3.25 & \textbf{4.01} & 2.76 & 3.54 & 3.66 & 3.78 & 3.78 & 3.31 & \textbf{3.99} & 1.41 & 3.24 & 3.28 & 3.58 & 3.62 & 3.14 & \textbf{3.99}  \\
 & ESTOI & 0.914 & 0.869 & \textbf{0.895} & 0.715 & 0.898 & \textbf{0.905} & 0.878 & 0.885 & 0.703 & 0.791 & 0.808 & 0.844 & \textbf{0.847} & 0.837 & 0.835 & 0.702 & 0.779 & 0.649 & \textbf{0.777} & 0.775 & 0.765 & 0.762 & 0.654 & 0.750 \\
 & ViSQOL & 4.636 & 4.246 & \textbf{4.596} & 3.967 & \textbf{4.410} & 4.384 & 4.199 & 4.298 & 3.534 & 3.847 & 3.476 & \textbf{3.765} & 3.615 & 3.648 & 3.633 & 3.265 & 3.628 & 2.611 & 3.301 & 3.164 & 3.194 & 3.150 & 2.930 & \textbf{3.411} \\
\midrule
\multirow{4}{*}{\rotatebox{90}{\textbf{Average}}} & \cellcolor{gray!20}TPR@1\%FPR & \cellcolor{gray!20}99.8 & \cellcolor{gray!20}43.2 & \cellcolor{gray!20}49.3 & \cellcolor{gray!20}\textbf{13.4} & \cellcolor{gray!20}98.0 & \cellcolor{gray!20}99.7 & \cellcolor{gray!20}92.8 & \cellcolor{gray!20}94.2 & \cellcolor{gray!20}4.8 & \cellcolor{gray!20}\textbf{1.8} & \cellcolor{gray!20}62.2 & \cellcolor{gray!20}48.3 & \cellcolor{gray!20}58.3 & \cellcolor{gray!20}23.9 & \cellcolor{gray!20}20.0 & \cellcolor{gray!20}4.2 & \cellcolor{gray!20}\textbf{1.8} & \cellcolor{gray!20}25.4 & \cellcolor{gray!20}12.8 & \cellcolor{gray!20}26.2 & \cellcolor{gray!20}15.5 & \cellcolor{gray!20}14.3 & \cellcolor{gray!20}1.5 & \cellcolor{gray!20}\textbf{1.2}  \\
 & UTMOS & 4.01 & 3.67 & \textbf{3.92} & 2.58 & 4.05 & 4.07 & \textbf{4.11} & 4.09 & 3.30 & 4.06 & 3.00 & 3.81 & 3.93 & 4.02 & 4.02 & 3.41 & \textbf{4.04} & 1.47 & 3.46 & 3.53 & 3.76 & 3.80 & 3.23 & \textbf{4.05}  \\
 & ESTOI & 0.977 & 0.928 & \textbf{0.958} & 0.749 & 0.963 & \textbf{0.968} & 0.939 & 0.947 & 0.736 & 0.813 & 0.871 & 0.903 & \textbf{0.909} & 0.897 & 0.894 & 0.730 & 0.801 & 0.697 & 0.822 & \textbf{0.825} & 0.810 & 0.806 & 0.679 & 0.774 \\
 & ViSQOL & 4.818 & 4.474 & \textbf{4.777} & 4.131 & \textbf{4.627} & 4.601 & 4.403 & 4.494 & 3.666 & 3.945 & 3.736 & \textbf{4.001} & 3.871 & 3.896 & 3.882 & 3.410 & 3.736 & 2.813 & 3.475 & 3.374 & 3.385 & 3.330 & 3.056 & \textbf{3.537} \\
\bottomrule
\end{tabular*}
\end{table*}

\afterpage{%
\begin{figure*}[t!]
\centering
\includegraphics[width=0.92\textwidth]{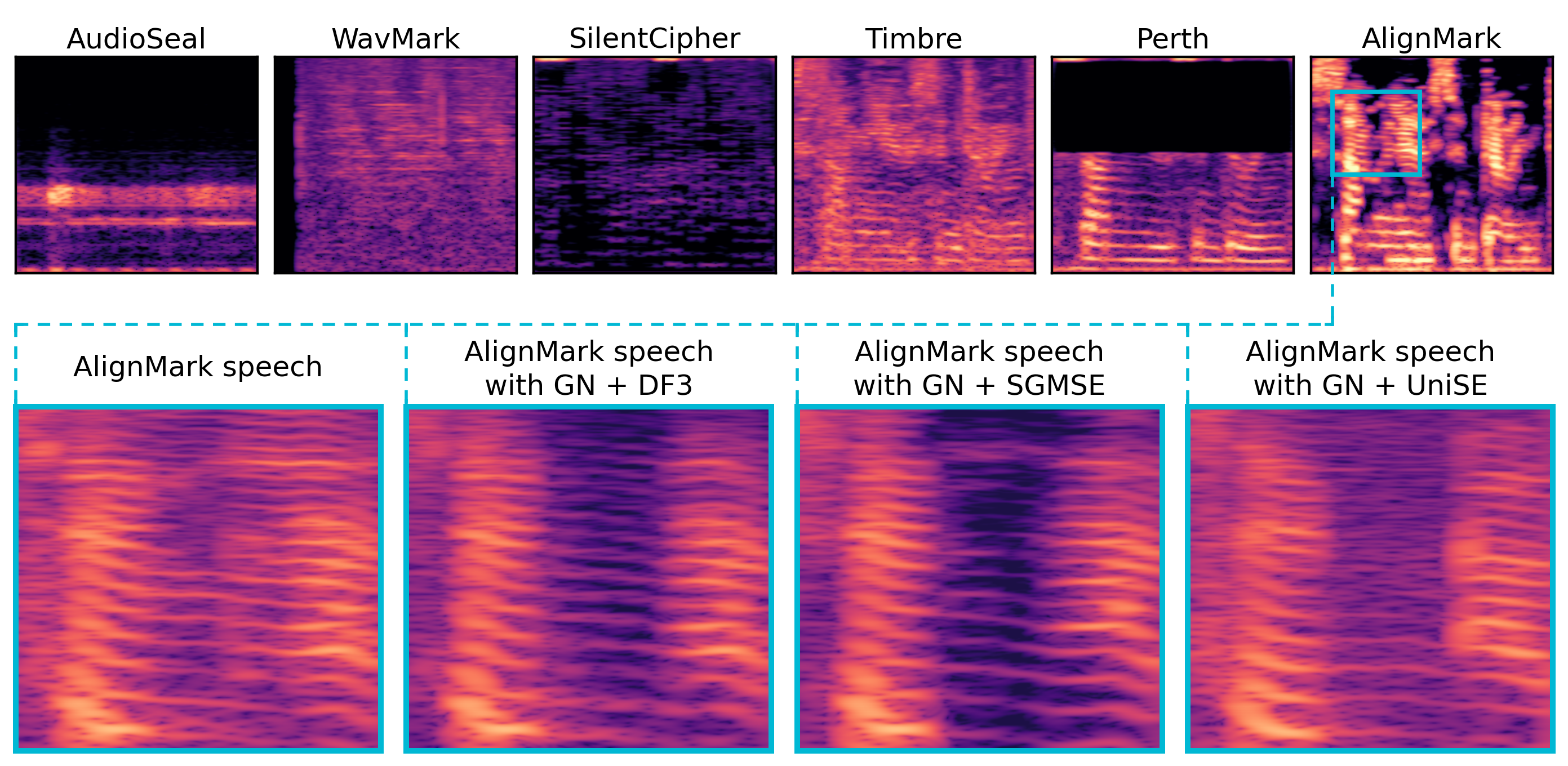}
\caption{Spectrogram visualization of neural watermarks under
speech enhancement. We visualize the watermark residual spectrogram
between the original and the watermarked (AudioSeal, WavMark,
SilentCipher, Timbre, Perth, AlignMark) speech on the top row.
Specifically, a region of the AlignMark residual that carries
harmonic information, marked by a blue rectangle, is observed further
in the bottom row, which shows the detailed spectrogram of the
AlignMark speech with the GN $10$\,dB $+$ SE attack pipeline.}
\label{fig:alignmark}
\end{figure*}}

\subsection{Results}
\label{ssec:results}
\rev{Table~\ref{tab:main} reports the results under four attack
conditions: no attack (WM only), neural re-synthesis (Re-synthesis),
SE without noise injection (SE only), and the GN $+$ SE attack
pipeline at $20$ and $10$\,dB.} \rev{Except for WM
only, the best value within each group is highlighted in bold.}

\textbf{WM only.}
All six watermarks are detected at a TPR of
$99.5$--$100.0\%$ with competitive perceptual metrics, \rev{whereas
AlignMark introduces a noticeably stronger perturbation than the others.}

\textbf{Re-synthesis.}
EnCodec at $24$\,kbps removes WavMark, SilentCipher and Timbre,
but leaves AudioSeal, Perth and AlignMark at a TPR of $100.0$,
$60.0$ and $97.5$. BigVGAN\,v2 removes AudioSeal, WavMark and
SilentCipher while attaining the best average perceptual
quality (UTMOS $3.92$, ESTOI \rev{$0.958$}, ViSQOL \rev{$4.777$}) among the
three baselines, yet preserves Timbre, Perth and AlignMark at a TPR
of $100.0$, $100.0$ and $91.0$. AudioLDM is the most destructive,
lowering the average TPR to $13.4$, but incurs the highest perceptual
cost (UTMOS $2.58$, ESTOI \rev{$0.749$}, ViSQOL \rev{$4.131$}).

\textbf{SE only.}
To ablate the role of the noise injection step, we apply each SE
model directly to the watermarked audio. The SE families split
sharply on the watermark. Demucs, \rev{DF3}, SGMSE, and
FlowSE all leave the watermark largely intact (average TPR
$98.0$, $99.7$, $94.2$ and $92.8$). GenSE and UniSE are
the exception: they almost fully remove every
watermark except AlignMark, which
retains $28.0\%$ and $9.5\%$ TPR respectively.
\rev{In terms of perceptual quality, discriminative SE and STSE outperform
LMSE, whose token-based regeneration discards more information from
the input speech. Among the LMSE models, UniSE scores higher than GenSE
on all three perceptual metrics, which could be attributed to its additional mel-feature
input that retains more fine-grained acoustic detail for speech
reconstruction.}

\textbf{GN $+$ SE.}
Pure Gaussian noise has limited destructive power against \rev{neural}
watermarks. At \rev{GN} $20$\,dB, all watermarks except
SilentCipher \rev{remain detectable}: the TPR of
AudioSeal, WavMark, Timbre, Perth and AlignMark are $60.0\%$,
$28.0\%$, $98.5\%$, $86.5\%$ and $99.5\%$, respectively. At
\rev{GN} $10$\,dB, Gaussian noise almost removes every watermark
except Timbre ($46.5\%$) and AlignMark ($98.5\%$), while degrading
perceptual quality (average UTMOS $1.47$) well below the level
acceptable for normal communication. We next cascade a pretrained SE
model after the noise injection, both to recover quality and to
observe whether the SE removes the watermark further or restores it.

\textbf{Gaussian noise followed by discriminative SE cannot fully
remove the watermark.} \rev{At GN $20$\,dB, Demucs attains the best
average ViSQOL ($4.001$). Its TPRs on WavMark and SilentCipher are $6.0$ and
$1.5$, while those on Timbre, Perth and AlignMark drop to $91.0$,
$54.5$ and $73.0$, respectively. DF3 attains the highest average ESTOI ($0.909$) and reduces
the TPRs of WavMark and SilentCipher to $9.5$ and $0.5$, but fails to
remove Timbre, Perth and AlignMark ($86.0$, $87.0$ and $99.5$). Both discriminative SE models slightly restore
the TPR of AudioSeal, and exhibit
the same trend in watermark removal across different GN levels.}

\textbf{Gaussian noise combined with generative SE collapses the
watermark.} At \rev{GN} $20$\,dB, STSE (SGMSE and FlowSE) drives
AudioSeal, WavMark, SilentCipher and Perth to TPR\,$\leq 4\%$ and
the more resilient Timbre to \rev{$19.0$ and $37.0$, respectively}; at \rev{GN} $10$\,dB
all five watermarks above are almost completely removed.
AlignMark is the single
exception, which STSE is almost unable to remove: the
watermark energy it aligns
with the speech feature distribution is redrawn together with the
harmonics it rides on, so a spectrogram-domain reconstruction
reproduces it. Only LMSE breaks it --- GenSE and UniSE take
AlignMark to TPR $23.5\%$ and $9.5\%$ at \rev{GN} $20$\,dB and to
$8.0\%$ and $6.5\%$ at \rev{GN} $10$\,dB, while holding the other
five watermarks at or below $1.5\%$. \rev{Notably, UniSE strikes a
favorable balance between watermark removal and perceptual quality,
with an average UTMOS of $4.05$ and ViSQOL of $3.537$ at GN
$10$\,dB.}

\subsection{\rev{Vulnerability Analysis and Visualization}}
\label{ssec:mechanism}
\rev{To analyze the vulnerability of the watermarks, we visualize them
in Fig.~\ref{fig:alignmark}. The top row shows the residual
spectrogram between the original and the watermarked speech for each
of the six watermarks. The bottom row zooms into the region marked by
the blue rectangle and compares the AlignMark speech with the outputs
of three representative GN $+$ SE attack pipelines (DF3, SGMSE and
UniSE) at GN $10$\,dB. These visualizations suggest that the
difficulty of removing a watermark depends on the distribution of the
watermark residual and the operating mechanism of the SE model.}

\textbf{Discriminative SE\rev{.}} It produces a deterministic estimate of
the clean signal, either as
direct waveform regression (Demucs) or as a spectral mask
(\rev{DF3}). The distribution and strength of watermark
perturbations differ markedly from those of the additive noise used
in SE model training. As a result, the model has no prior knowledge
of such perturbations, allowing most of the watermark to survive the
GN $+$ SE pipeline. Fig.~\ref{fig:alignmark} shows that DF3 tends
to preserve the continuous harmonic structure of the input and
denoises less aggressively than the generative models, so the
AlignMark signature survives.

\textbf{Generative SE\rev{.}} When STSE (SGMSE and FlowSE) is applied
directly to the watermarked speech, it is close to an identity
mapping and therefore cannot remove the watermark. Injecting Gaussian noise
pushes STSE into a regenerative regime, where the entire spectrogram
is reconstructed from the learned clean-speech prior.
We attribute the removal of the watermark by the GN $+$ STSE
pipeline to two mechanisms. \rev{First, and
most importantly, the more robust watermarks, Timbre and Perth,
concentrate their residual near regions of high spectral energy, and
GN $+$ STSE redraws exactly these regions: voiced harmonics and
formant continuity are resampled from that prior. The watermark,
which lives in the regions the model just redrew, loses its
structural integrity.} Second, the portion of the watermark
residual whose energy distribution overlaps with the additive
Gaussian noise is removed during the generative denoising step. However,
AlignMark embeds with higher energy and closer alignment to the
harmonic structure than the other watermarks~\cite{li2026alignmark},
so the partial spectrogram reconstruction performed by STSE is not
sufficient to destroy it.
Fig.~\ref{fig:alignmark} shows that, compared with the
discriminative pipeline, GN $+$ SGMSE denoises more aggressively and
leaves visible breakpoints in the formant transition bands of the
AlignMark-watermarked speech.

In LMSE (GenSE and UniSE), speech is represented as a sequence of
discrete tokens produced by an autoregressive language model, and the
waveform is finally reconstructed from them by a codec decoder.
Those tokens are optimized to carry high-dimensional semantic and
acoustic information, not to preserve fine-grained waveform detail,
so a watermark residual could be preserved only to the extent that
it is encoded into the latent space preceding token
quantization. The AlignMark speech with GN $+$ UniSE in
Fig.~\ref{fig:alignmark} shows a more pronounced change in harmonic
structure than under DF3 and SGMSE: the harmonic envelope becomes
more regular, and the formant transition lines at high frequency
are blurred. Since formant trajectories and fine harmonic structure
also carry speaker-specific cues, this regularization alters the
speaker timbre to some extent and lowers the speaker similarity of
the LMSE output to its input. \rev{We therefore argue that the most effective way to
remove a watermark is to alter the structure of the harmonic region
during denoising, as much as perceptual quality allows.}

\section{Conclusion}
\label{sec:conc}
In this paper we introduce and systematically analyze a strict
black-box watermark removal attack that cascades Gaussian noise with
six \rev{SE} models. By studying the performance of six
watermarking algorithms within our attack pipeline, we find that a
watermark is removed most effectively when the SE model both
denoises and regenerates the harmonic regions. \rev{On the whole, future robustness evaluation of audio
watermarking should address not only signal-level distortion and
neural re-synthesis but also speech enhancement, which this study
identifies as a practical vulnerability.}

\vfill\pagebreak
\balance
\bibliographystyle{IEEEbib}
\bibliography{refs}

\end{document}